\documentstyle[preprint,aps]{revtex}
\begin{document}
\draft

\title{$Z \rightarrow b {\bar b}$ in a composite model of fermions}
\author{Pham Quang Hung}
\address{Department of Physics, University of Virginia, Charlottesville,
Virginia 22901, USA}
\date{\today}
\maketitle
\begin{abstract}
A composite model of fermions is proposed to explain the "anomaly" in
$Z \rightarrow b {\bar b}$ and, to a lesser extent, 
in $Z \rightarrow c {\bar c}$. 
It contains a
{\em nonsequential} fourth family whose mass of one member
(the charge -1/3 quark) 
is constrained to be between 47 GeV and 49 GeV. The charge +2/3
quark is constrained to lie between 67 GeV and 107 GeV.
This opens up the exciting prospect for
near-future discoveries at LEP2 and possibly at the Tevatron.
\end{abstract}

\narrowtext

% body of paper here
\section{Introduction}

Precision tests of the Standard Model (SM) have reached a level where it
"might" now be possible to look for indirect evidence of new physics and/or
new degrees of freedom. One example  is the {\em apparent} discrepancy
between theory and experiment in the value of the ratio
$R_{b} \equiv \Gamma (Z \rightarrow b {\bar b})/ 
\Gamma (Z \rightarrow had )$ \cite{langacker}. This 
discrepancy which increases
with $m_t$, reaches the 3 $\sigma$ level when $m_t$ reaches 175 GeV.
In addition the ratio $R_{c} \equiv \Gamma (Z \rightarrow c {\bar c})/ 
\Gamma (Z \rightarrow had )$ is 2 $\sigma$ smaller than the SM
prediction. In addition there seems to be some discrepancy
between the measurements of the left-right asymmetry $A_{LR}$
done at SLD and at LEP.
If one also includes the {\em apparent} disagreement between the
QCD coupling $\alpha_{S}$ determined at "low" energy and evolved to
$M_Z$ with that determined by the Z-lineshape, one is tempted to
think that one might be already seeing some new kind of physics.
It is therefore very crucial 
to confirm or disprove these so-called discrepancies.
Let us nevertheless assume that they are not mere statistics
and examine what kind of new physics that can be possible and
what predictions that can be tested in the near future. 
Even if
the discrepancy were to disappear, this would put a severe constraint
on this type of new physics. 

In this manuscript, a mechanism is proposed to explain the apparent
increase of $R_b$ and, as a consequence, 
the decrease in $R_c$ , and to make further predictions
on other branching ratios, and ultimately to constrain
the new physics involved in the mechanism itself. 
It is based on the assumption
that there is a new, heavy, {\em nonsequential} down
quark ($Q = -1/3$) ( part of a new family) with mass greater than 47 GeV
and whose $q {\bar q}$ bound state(s) (by QCD) mixes with the
Z boson. By {\em nonsequential}, 
we mean that the fermions of
the new family has very little
mass mixing with fermions of the other
three generations.
(The description of a concrete model is given below.)
Consequently,
the following predictions are made for the hadronic widths.
We make the following predictions.
There is a {\em decrease} in $R_{c} \equiv \Gamma (Z \rightarrow c {\bar c})/ 
\Gamma (Z \rightarrow had )$ and
$R_{u} \equiv \Gamma (Z \rightarrow u {\bar u})/ 
\Gamma (Z \rightarrow had )$ ,
and an {\em increase} in $R_{d} \equiv \Gamma (Z \rightarrow d {\bar d})/ 
\Gamma (Z \rightarrow had )$,
$R_{s} \equiv \Gamma (Z \rightarrow s {\bar s})/ 
\Gamma (Z \rightarrow had )$ and
$R_{b} \equiv \Gamma (Z \rightarrow b {\bar b})/ 
\Gamma (Z \rightarrow had )$, 
all in comparison with the SM predictions. All of these changes are 
predicted in terms of a single increase in $\Gamma (Z \rightarrow b {\bar b})$.
If the new, heavy vector meson couples universally (with a different
strength in principle) to the ordinary leptons
then $\Gamma (Z \rightarrow \nu {\bar \nu})$
and $\Gamma (Z \rightarrow l^{+} l^{-})$ are also predicted to
decrease and increase respectively.
Our predictions are in basic agreement with all Z-pole observables
except for one: the left-right asymmetry $A_{LR}$. There our
prediction is in agreement with the SLD data. This is perhaps
also an indication of new physics such as the type discussed
in this paper. In this regard, it is important to stress the
fact that one has to take into account, in {\em any} discussion
of new physics affecting $R_b$, other electroweak observables
as well, such as $A_{LR}$, $\Gamma_Z$, $\sigma_{had}$, etc...,
and not just $R_b$ and $R_c$.

Some comments will be made regarding the possible mass ranges of
the new fermions as well as the range of compositeness scales
of the model to be described below.

\section{A Model}

In this section we shall describe a model 
which motivates the subsequent phenomenological
discussion. We shall expose mainly the salient features of the model needed
for this discussion,
leaving out some details for a subsequent paper which will focus on
the construction of the model and its implications concerning mass matrices.

The model we are concerned with in this paper is a confining model
in the manner of Abbott-Farhi \cite{farhi}, 
where the usual quarks and leptons are
viewed as composites of more fundamental fermions and scalars. In
contrast with the Abbott-Farhi model where the confining gauge group is
the electroweak group, here it is a {\em family} gauge group
which is confining. Also, in contrast with the composite
models constructed long ago by various authors, here W and Z {\em are}
fundamental gauge fields while there exists composite (global) family
vector bosons with masses as high as the compositeness scale itself.
To summarize, the Abbott-Farhi model contains composite weak vector bosons
while the model presented here contains composite horizontal
or family vector bosons.
The reason for considering this kind of model
is a desire to understand the family structure of the standard
model and its mass matrices.

The model is a Left-Right symmetric extension of the Standard
Model with a confining Left-Right horizontal gauge group. The
gauge structure is $\{SU(3)\otimes SU(2)_L \otimes SU(2)_R
\otimes U(1)_{B-L}\} \otimes SU(2)_{HL} \otimes SU(2)_{HR}$,
with $SU(3)$ being the usual color gauge group. 

Let us recall
that in Abbott-Farhi-type models, the scalar sector has an additional
global $SU(2)$ and it was this $SU(2)$ that acted as an effective
weak interaction group. Let us also recall that there the preonic 
fermions and scalars transform as singlets and doublets under
that global symmetry respectively. (As a result, quarks and
leptons which are fermion-scalar bound states and W and Z which
are scalar-scalar bound states transform as doublets and triplets
under the global $SU(2)$ symmetry respectively.) What are the
differences between the present model and the Abbott-Farhi one?

Here the additional global symmetries will be 
$SU(2)_{GL} \otimes SU(2)_{GR}$
in analogy with the Abbott-Farhi model, with the {\em difference} being
that these global symmetries are now attached to {\em horizontal }left
and right symmetries. The minimal preonic particle content is given by:
$\Psi_{qL} = (3,2,1,2,1,2,1,1/3)$;
$\Psi_{qR} = (3,1,2,1,2,1,2,1/3)$;
$\Psi_{lL} = (1,2,1,2,1,2,1,-1)$;
$\Psi_{lR} = (1,1,2,1,2,1,2,-1)$;
$\phi_{L} = (1,1,1,2,1,2,1,0)$;
$\phi_{R} = (1,1,1,1,2,1,2,0)$, where $\Psi$ and $\phi$ denote fermions
and scalars respectively. The transformations are with respect to
$SU(3)\otimes SU(2)_L \otimes SU(2)_R
\otimes SU(2)_{HL} \otimes SU(2)_{HR} \otimes
SU(2)_{GL} \otimes SU(2)_{GR} \otimes U(1)_{B-L}$. 
Notice that in our minimal model the scalar
fields are singlets with respect to the electroweak group.

Let us assume that $SU(2)_{HL} \otimes SU(2)_{HR}$ is confining. The
physical quarks and leptons, which are now composite objects, transform as:
$q_L = (\Psi_{qL} \phi_{L}) = (3,2,1,1,1,1+3,1,1/3)$;
$q_R = (\Psi_{qR} \phi_{R}) = (3,1,2,1,1,1,1+3,1/3)$;
$l_L = (\Psi_{lL} \phi_{L}) = (1,2,1,1,1,1+3,1,-1)$;
$l_R = (\Psi_{lR} \phi_{R}) = (1,1,2,1,1,1,1+3,-1)$.
Notice that under the global horizontal (family) group
$SU(2)_{GL} \otimes SU(2)_{GR}$, the left and right-handed quarks and
leptons transform as a triplet plus a singlet, i.e. there are four
families in this model, with the fourth one (singlet) being
separate from the other three in the lowest order. This is the statement
made in the introduction. 

A remark is in order here. If the preonic
quarks and leptons were to transform as singlets under the global
horizontal group, there would only be two families of {\em composite}
quarks and leptons. To incorporate the third family, one would
have to add another set of preons with the result that one now
has two sets of disjointed double families. This does not appear to be
the case in reality and, in any case, one also ends up with four families.
The previous scenario of three connected families and one disjointed
family (in the lowest order) seems to be more desirable.

We would like to make one more remark. Another possible scenario
not considered here is to keep $SU(2)_{HL} \otimes SU(2)_{HR}$
unconfined and to endow the preonic fermions and scalars with
some extra confining gauge symmetry and that they transform as
fundamentals under that extra gauge symmetry. Again, the (fermion-
scalar) composites would decompose into triplets and singlets
of the (now gauged) horizontal symmetry.

The moral of the story is that as long as the (gauge or global)
horizontal symmetry is $SU(2)$ and that the {\em preonic} fields
are doublets, one would get three connected families (the
standard three families) and one
disconnected one (in the lowest order) at the composite level.
Let us denote this {\em nonsequential} family by $Q = ({\cal R},
{\cal P})$ for the quarks and by $L = ({\cal N}, {\cal E})$ for
the leptons. This nonsequential family behaves {\em exactly} like
the standard three families under the gauge group
$SU(3)\otimes SU(2)_L \otimes SU(2)_R \otimes U(1)_{B-L}$. We would
like to stress this point in order to avoid any misunderstanding:
the nonsequential family is just another generation which is
disconnected from the other three (in the lowest order).

Below the scale of "compositeness", there can be, besides the
usual gauge interactions among the composite fermions,
several four-fermi interactions, some of which are relevant for the
present discussion and some for the study of mass matrices. (They
can be viewed as resulting from the exchange of some composite
bosons.) We are mainly concerned here with 
the interactions between the nonsequential
fourth generation and the other three. This is because we
are interested in the effects of the nonsequential
generation on physics involving ordinary quarks and
leptons. To this end, let us
denote $G_{gauge} = SU(3)_c \otimes SU(2)_L \otimes
SU(2)_R \otimes U(1)_{B-L}$ and $G_{global} = SU(2)_{GL}
\otimes SU(2)_{GR}$. The interactions should be invariant
under $G_{gauge}$ but not necessarily under $G_{global}$
which could be broken explicitely by these interactions.
Let us recall that the nonsequential fourth generation
is singlet under $G_{global}$.

There are several scenarios. We shall present one of such
scenarios here. Let us assume for example that there is
a neutral interaction between the nonsequential
fourth generation and the other three of the form which
is $G_{global}$-invariant, namely 
\begin{equation}
{\cal L}_{q0} = (g_{q}^2/\Lambda^2) \sum_i
{\bar {\cal Q}_i} \gamma_{\mu} {\cal Q}_i  \sum_j 
{\bar q_j} \gamma^{\mu} (1- \gamma_5)q_j, 
\label{LqO}
\end{equation}
where the sums over $i$ and $j$
refer to all quarks of the new and "old"
generations respectively. ${\cal L}_{q0}$ will provide the kind
of coupling which is used here and whose phenomenological
implications concerning $Z \rightarrow b {\bar b}$ are discussed below.
In addition, we could have the following interactions among the
"new" quarks and the "old" leptons:
\begin{equation}
{\cal L}_{l0} = (g_{l}^2/\Lambda^2) \sum_i
{\bar {\cal Q}_i} \gamma_{\mu} {\cal Q}_i  \sum_j 
{\bar l_j} \gamma^{\mu}(1- \gamma_5)l_j, 
\label{LlO}
\end{equation}
where, in principle, $g_l \neq g_q$. These two equations represent
the relevant interactions for describing the phenomenology of
the new, heavy quark bound state mentioned earlier and to which we
shall come back below. We then discuss the limitation of these
assumptions and suggest possible modifications.

In addition we shall assume the
following $G_{global}$ breaking term:
\begin{equation}
{\cal L}_B = (g_{b}^2/\Lambda^2) (
{\bar {\cal Q}} \Gamma \frac{\vec{\tau}}{2} {\cal Q} +
{\bar {\cal L}} \Gamma \frac{\vec{\tau}}{2}{\cal L}) \cdot
{\bar l}_{3} \Gamma \frac{\vec{\tau}}{2} l_{3}, 
\label{LB}
\end{equation}
where $l_3 = (\nu_{\tau}, \tau)$, and $\Gamma =
1,\gamma_{\mu} (1-\gamma_5), (1-\gamma_5)$, etc.... 
${\cal L}_B$ is $G_{global}$-
breaking because only $l_3$ is present. At present, we
have not explored the possible sources for this term. One
possibility would be the mixing of the "charged Higgs"
coupled to the nonsequential family with the
corresponding one which couples to the standard families. (In
our scenario, it is unavoidable to have several physical scalars.)
Since the "standard family" charged Higgs will couple preferentially to
$\tau \nu$ as far as the lepton sector is concerned, it might be
possible that the couplings and masses (of the mixed one) are
such as to favor ${\cal R} \rightarrow {\cal P} \tau^{+}
\nu$. It will be seen at the
end of the paper that this kind of interaction which provides
a non-standard decay mode for the ${\cal R}$ quark is severely
constrained by CDF and D0. Another
remark is in order here. The $\Lambda$'s in ${\cal L}_{q0}$ and
${\cal L}_B$ are not necessarily the same. For simplicity
we shall take them to be equal to each other, keeping in
mind that they can differ in value.

${\cal L}_{q0, l0}$ will form the backbone of the phenomenology of this
paper while ${\cal L}_B$ will be seen to provide the dominant leptonic 
decay mode of the fourth generation provided
$g_{b}^2/\Lambda^2$ is large enough which we will see to be the case.
This will provide the rationale for its unobservability at the present
time because the leptonic decay of ${\cal R}$ will be mostly 
into ${\cal P} \tau \nu$. We shall come back to this point below.

One last remark is in order. In general, one expects all kinds of
four-fermi interactions, including two classes which are not directly
relevant to the present discussion. One of such classes is the 
four-fermi interactions involving only the fourth generation. For
obvious reasons we are not interested in such a class in this paper.
The other one is the four-fermi interactions involving only
fermions of the first three generations. These are the kind of
interactions that we shall use to construct mass matrices in 
a separate paper. The nature of these interactions, including
the coefficients $g_i^2/\Lambda_i^2$ in front, is however
unknown. Each assumption concerning one of these interactions
will have some phenomenological consequences which results
in the experimental constraints on $g_i^2/\Lambda_i^2$ which
are {\em not necessarily} the same as those given in
Eq.\ (\ref{LqO}, \ref{LB}). To be more complete here we shall
write down a generic term of the form
\begin{equation}
{\cal L}_f = (g_i^2/\Lambda_i^2) {\bar f}_1 \Gamma f_2
{\bar f}_3 \Gamma^{'} f_4,
\label{Lf}
\end{equation}
where $f$ denotes some generic third-generation fermion, $\Gamma$
denotes some generic Lorentz and internal symmetry structure, 
and the subscript $i$ labels the coefficients which appear
in front of these interactions.

We shall now come to the main part of this paper, namely the effect
of the nonsequential fourth generation, specifically the quarks,
on the decay of the Z boson.

\section{Phenomenological Analysis of $Z \rightarrow \lowercase{b} 
\lowercase{{\bar b}}$}

Although the discussion presented below concerning the
decay mode $Z \rightarrow b {\bar b}$ is related to our
composite model, we shall present it in a way which is
general enough to be applicable to other models as well. The
only assumption is the existence of nonsequential fourth family
with a particular coupling to the other three families.

Before we start the discussion on the effects of this
nonsequential fourth family on $Z \rightarrow b {\bar b}$,
a few remarks are in order concerning a potential mixing
between the SM Z boson and $Z^{\prime}$ coming from
$SU(2)_L \otimes SU(2)_R \otimes U(1)_{B-L}$. How big or
how small such a mixing is depends on the details of the
Higgs sector. We shall {\em assume} that such a mixing, if
it exists, is small enough  as to give a negligible
contribution to $Z \rightarrow b {\bar b}$ and other 
observables. In fact, an analysis of precision electroweak
data as applied to extended gauge models, in particular
a Left-Right model as used in this paper, by Ref. 
\cite{altarelli,luo} constrained the mixing to be very small.
By parametrizing the mixing in terms of an angle $\xi$,
namely $Z= cos\xi Z_S + sin\xi Z_N$ and
$Z^{\prime} = -sin\xi Z_S + cos\xi Z_N$, where $Z_S$ and
$Z_N$ are the SM and new gauge bosons before mixing, the
authors of Ref. \cite{altarelli,luo} found that $\xi$ is
constrained by precision electroweak data to be less than
1 \%. The reader is referred to Ref. \cite{altarelli,luo}
for more details. 

From hereon we shall assume that the
mixing with $Z^{\prime}$ is {\em negligible}. The mixing is assumed 
to be negligible both at the tree level and even at the
one-loop level (through the top quark for example) if 
$Z^{\prime}$ is heavy enough. Although the possibility
of various deviations which might come from the mixing with
$Z^{\prime}$ is interesting in its own right, we would
like to present yet another mechanism for such a deviation
and choose to neglect the effect of $Z^{\prime}$ if the
mixing is assumed to be very small.
We shall therefore
concentrate on the effects of the mixing between $Z$ and
a heavy quarkonium.

As we have mentioned above, let 
us denote this {\em nonsequential} family by ($\cal R$,
$\cal P$) for the quarks and by ($\cal N$, $\cal E$) for
the leptons. For reasons to be given below let us assume
that the ($Q=-1/3$) quark has a mass $m_{\cal P} < m_{\cal R}$.
We also assume that the up-type quark ${\cal R}$ is heavy
enough so that ${\cal R}{\bar{\cal R}}$ QCD bound states are
well above the ${\cal P}{\bar{\cal P}}$ open threshold.
The ${\cal P} {\bar{\cal P}}$ QCD bound states can be described
by Richardson's potential. Such an analysis has been carried out
long ago by \cite{gilman} for the $ ^3S_1$ $t {\bar t}$
bound states, but unfortunately in the now-obsolete range of
$m_t \sim 40-50$ GeV. This analysis can however be applied to
any quark in a similar mass range or higher, especially
for our case where $m_{{\cal P}} > 46$ GeV.
(The mass shift of the Z boson due to this mixing is negligible
\cite{gilman}.)

${\cal P} {\bar {\cal P}}$ QCD bound states which can mix with Z 
are either vector, axial vector, or both. In what follows
we shall neglect the mixing of Z with the axial vector states
since it goes like $\beta^3$ \cite{zerwas,gilman}. Consequently
we shall focus only on the vector meson ($^3S_1$) bound states.
In particular, we shall first examine the mixing of the ground
state $1S$ with Z. In the mass range considered here, the ground
state $1S$ is sufficiently far from open-$\cal P$ threshold so
that  the mass-mixing formalism can be applied. Denoting the
$1S$ ($J^{PC} = 1^{--}$) state by $V^0$, the result of $V^0$
and $Z^0$ mixing is given in terms of the mass eigenstates \cite{gilman}
\begin{mathletters}
\begin{equation}
|V\rangle = cos\frac{\theta}{2} |V_0\rangle - sin\frac{\theta}{2}
|Z_0\rangle,
\label{massvec1} 
\end{equation}
\begin{equation}
|Z\rangle = sin\frac{\theta}{2} |V_0\rangle + cos\frac{\theta}{2}
|Z_0\rangle,
\label{massvec2} 
\end{equation}
\end{mathletters}
for the mass eigenvectors and where
\begin{equation}
\theta = sin^{-1} (\delta m^2 / \Delta^2),
\label{angle1}
\end{equation}
with
\begin{equation}
\Delta^2 = [\frac{(
M_{V_0}^2 - i \Gamma_{V_0} M_{V_0} - M_{Z_0}^2 + i \Gamma_{Z_0} M_{Z_0})^2}
{4}+ (\delta m^2)^2 ]^{1/2}.
\end{equation}
$\delta m^2$ is the off-diagonal element of the mass mixing matrix
and is given by \cite{gilman}
\begin{equation}
\delta m^2 = F_V [(\frac{g}{cos\theta_W})\frac{\frac{4}{3}sin^2\theta_W
-1}{4}],
\label{offdiag}
\end{equation}
where
\begin{equation}
F_V = 2 \sqrt{3} |\Psi (0)| \sqrt{M_{V_0}}.
\label{wfunc}
\end{equation}
and where the factor 3 comes from the number of colors and 
$|\Psi (0)|$ is the wave function at the origin which can be
computed using the Richardson's potential in QCD. 
The term inside the square brackets represents the {\em vector}
coupling of the $\cal P$ quark to the Z boson.

Let us assume that $M_V > M_Z$ and since present experiments are carried
out on the Z resonance, we need only to look at Eq.\ (\ref{massvec2})
to see how the presence of $V_0$ modifies the coupling of Z to
"light" quarks and leptons. This, as we claim in this manuscript,
is a possible source for the discrepancy seen in $\Gamma(b {\bar b})$.
From Eq.\ (\ref{massvec2}), one finds the physical Z couplings to
a given fermion $f$ to be
\begin{equation}
g_{Z f {\bar f}}^{V,A} = sin\frac{\theta}{2} g_{V_0 f {\bar f}}^{V,A}
+ cos\frac{\theta}{2} g_{Z_0 f {\bar f}}^{V,A},
\label{coupling}
\end{equation}
where $V$ and $A$ stand for vector and axial-vector couplings respectively.
$g_{V_0 f {\bar f}}^{V,A}$ and $g_{Z_0 f {\bar f}}^{V,A}$ are the
couplings before mixing.

Before mixing, the heavy quarkonium $V_0$ 
can decay into $f \bar{f}$ via $\gamma$
if there were no new physics involved. (The reader is referred to
Ref. \cite{barger} for a pedagogical discussion of this point.)
This source alone however gives only a small change to $R_b$.
A new and unconventional coupling
of $\cal P$ to $b$ ( and to other normal fermions as well)
is needed to bring $R_b$ closer to its experimental
value. We have seen in the previous section how
such coupling can arise in our composite model. Let us write
\begin{equation}
g_{V_0 q {\bar q}}^{V,A} = F_V G_{q}^{V,A}(s=M_Z^2) +
g_{new, q}^{V,A},
\label{coupling2}
\end{equation}
where $G_{q}^{V,A}$ can be found in\cite{barger}.
\begin{mathletters}
\begin{equation}
G_{q}^{V}(M_Z^2) = e^2 \frac{Q_q Q_{\cal P}}{M_Z^2} 
\label{GV}
\end{equation}
\begin{equation}
G_{q}^{A}(M_Z^2) = 0
\label{GA}
\end{equation}
\end{mathletters}
and where $Q_q$ and $Q_{\cal P}( = -1/3)$
are the electric charges. 
$g_{new,q}^{V,A}$ is the coupling of
$V_0$ to a quark $q$ and is found to arise from 
${\cal L}_0$ as we shall see below. 
We would like to constrain $g_{new,q}^{V,A}$
using the experimental value of $R_b$. A similar
term can be written for the coupling of $V_0$
to a lepton $l$ where one now has $Q_l$ and
$g_{new,l}$.

For the mass range considered below ,
namely $m_{\cal P} > 46 GeV$, $|\Psi (0)|$ is
such that \cite{gilman} $|\delta m^2| \ll |
M_{V_0}^2 - i \Gamma_{V_0} M_{V_0} - M_{Z_0}^2 + 
i \Gamma_{Z_0} M_{Z_0})^2|/2$ and consequently
\begin{equation}
sin\frac{\theta}{2}\approx \frac{\delta m^2}{M_{V_0}^2 
- M_{Z_0}^2 + i (\Gamma_{Z_0} M_{Z_0} - \Gamma_{V_0} M_{V_0})},
\label{sint}
\end{equation}
with $cos\frac{\theta}{2} \approx 1$.
Typically, $\theta/2 \approx 2-3 \times 10^{-2}$ and the deviation
of $cos\frac{\theta}{2}$ from unity will be of order $10^{-4}$ and
can be neglected considering the present level of precision.

The modified couplings of Z to a quark $q$ are now
\begin{mathletters}
\begin{equation}
\tilde{g}_{q}^V = (1+ \eta_{q,W}^V + \eta_{q,new}^V) g_{q}^V,
\label{gvtil}
\end{equation}
\begin{equation}
\tilde{g}_{q}^A = (1 + \eta_{q,new}^A) g_{q}^A,
\label{gatil}
\end{equation}
\end{mathletters}
where $W$ stands for electroweak and the $\eta$'s are complex numbers
and are defined by
\begin{mathletters}
\begin{equation}
\eta_{q,W}^{V,A}  = sin\frac{\theta}{2} F_V G_{q}^{V,A}(s=M_Z^2)/g_{q}^{V,A},
\label{etawre}
\end{equation}
\begin{equation}
\eta_{q,new}^{V,A}  = sin\frac{\theta}{2} g_{new,q}^{V,A}/g_{q}^{V,A},
\label{etanre}
\end{equation}
\end{mathletters}
where the explicit forms for $\eta_{q,W}^{V}$ and $\eta_{q,new}^{V,A}$
can be obtained by using Eqs.\ (\ref{coupling2},\ref{sint}).
{\em For simplicity}, we shall assume the new interactions to
be V-A, namely
$g_{new,q}^{V}$ = -$g_{new,q}^{A}$ = $g_{new,q}$.
This is consistent with ${\cal L}_{q0}$ in Eq. (1).
There the V-A nature of the "standard" quark (denoted by
q) current was explicitely assumed. 
We shall try to relate $g_{new,q}$ to the compositeness scale below.
Let us however be slightly more general and take $g_{new,q}$, 
for the moment, to
simply parametrize the
"new physics" involved in $Z \rightarrow b {\bar b}$ and 
extract it from $R_b$.

The modified coupling of $Z$ to a lepton $l$ can be written in a similar
fashion to Eqs.\ (\ref{gvtil}, \ref{gatil}) with the substitution
$q \leftrightarrow l$. In terms of the new physics, we now have
two parameters: $g_{new,q}$ and $g_{new,l}$. In principle, they
can be very different from each other.

In computing the Z widths using Eqs.\ (\ref{gvtil},\ref{gatil}) and the
range of mass mentioned earlier, one can safely {\em neglect}
terms proportional to $(Re\:\eta)^2$ and $(Im\:\eta)^2$ since
they turn out to be {\em at least} two orders of magnitude smaller
than terms proportional to $Re\:\eta$ (assuming $g_{new,f}^{V} < 1$).
(Considering the present level of precision, their inclusion is
irrelevant to the present discussion.) With this remark in mind,
the decay width for $Z \rightarrow f {\bar f}$ is now given by
\begin{equation}
\Gamma(Z \rightarrow f {\bar f}) = \Gamma_{f}^{SM} (1 + 
\delta_{new}^{f}),
\label{gam}
\end{equation}
where $f = q, l$ and where
\begin{equation}
\delta_{new}^{f} = \frac{2 (g_{f}^{V})^2 ( Re\:\eta_{W}^V +
Re\:\eta_{new}^{V})+ (g_{f}^{A})^2\:Re\:\eta_{new}^{A}) }
{(g_{f}^{V})^2 + (g_{f}^{A})^2}.
\label{delnew}
\end{equation}
In Eq.\ (\ref{gam}), $\Gamma_{f}^{SM}$ contains various radiative 
correction factors as well as mass factors such as defined in Ref.\
(\cite{Bernabeu}). We find
\begin{mathletters}
\begin{eqnarray}
\Gamma(had)& = &\Gamma^{SM}(had) + \delta_{new}^{u} ( \Gamma_{u}^{SM}
+ \Gamma_{c}^{SM}) \nonumber \\
& & \mbox{}+ \delta_{new}^{d} (\Gamma_{d}^{SM}
+ \Gamma_{s}^{SM} + \Gamma_{b}^{SM}),
\label{gamhad}
\end{eqnarray}
\begin{equation}
R_{f} =  \frac{R_{f}^{SM}(1+ \delta_{new}^f)}{1 +
\delta_{new}^{u} ( R_{u}^{SM}
+ R_{c}^{SM}) + \delta_{new}^{d} (R_{d}^{SM}
+ R_{s}^{SM} + R_{b}^{SM})},
\label{rhad}
\end{equation}
\end{mathletters}
where $R_{f} \equiv \Gamma(Z \rightarrow q_f {\bar q}_f) / \Gamma(had)$.
The central theme of this paper is the use of $R_b$ to obtain
information on the model proposed here. By using Eq.\ (\ref{rhad})
for $R_b$, one can extract the parameters $Re\eta_{b,new}^{V,A}$ and
consequently the {\em common} parameter $sin\frac{\theta}{2}g_{new}$
as a function of $M_{V_0}$. This will then be used to make predictions
on various ratios mentioned above and also on the total Z width. 
Finally $Re\eta_{new}^{V,A}$ will also give information
on the possible values for $g_{new,q}$ and consequently on the scales
of new physics as we shall see below.

We shall use the following experimental ratio \cite{langacker}: 
$R_b = 0.2219 \pm 0.0017$. In our analysis, the
SM predictions as functions of the top quark and 
Higgs masses (see e.g. \cite{Bernabeu}) are listed in Table 1. 
(Notice that the results of \cite{Bernabeu} are obtained
for $\alpha_S (M_Z) = 0.12 \pm 0.01$).

Our strategy is to extract $g_{new,q}$ from $R_b$ and to use it
to make predictions on $R_c$ and $R_s$ ($R_u$ and $R_d$ are practically
the same as these two respectively). They are listed
in Table 2. To make predictions concerning
the leptonic sector, one has to know $g_{new,l}$. This can be done
by choosing values that fit
$R_e \equiv \Gamma(had) / \Gamma(e {\bar e})$ and, consequently,
use them to predict
$A_{LR}$, $\sigma_{had}$, and $\Gamma_Z$. This is the procedure we
choose to follow in this paper. The results 
are listed in Table 3 along with the respective experimental values.
Let us now discuss these results.

A look at Table 2 shows that our predictions for $R_c$ are in
basic agreement with the experimental value. The basic observation here
is there is a {\em decrease} in $R_c$ with respect to the SM prediction
which is shown in Table 1. In our model this decrease is {\em real}
and is due to an increase in $R_b$. The amount of 
the decrease in $R_c$, for
a given top quark mass, is entirely {\em determined} by the amount
of increase in $R_b$. This prediction is fixed in our model.

We also predict an increase in $R_s$ =  $R_d$, and a decrease in
$R_u$ = $R_c$. The results are shown in Table 2. These predictions
are insensitive to the Higgs mass.

Notice that
an increase in the ratio for a down-type quark corresponds to
a decrease in the ratio for an up-type quark and vice versa.
This happens because $Re\eta_{f,new}^{V,A}$ is positive for
$f=u,c$ and negative for $f=d,s,b$. ($V_0$ is a ${\cal P}
{\bar{\cal P}}$ bound state.) 
Also notice that, in terms of the experimental $R_b$, one
can also write
$R_{c,s} = (\Gamma_{c,s,u}^{SM} / \Gamma_{b}^{SM})((1+
\delta_{new}^{c,s})/(1+\delta_{new}^{b}))R_b$. It turns
out that $(1+\delta_{new}^{c,s})/(1+\delta_{new}^{b})$, and
hence $R_{c,s}$, is
independent of $M_{V}$.

Beside $R_b$ and $R_c$, can the presence of $V$ and its
mixing with $Z$ affect other observables such the Z-width,
$\sigma_{had}$, and asymmetries such as $A_{LR}$? In
particular, will these observables deviate significantly from
their standard model predictions and hence signal the
presence of $V$ even if one is a few V-width away from its peak?
These are the questions which we will address below.

As we have discussed above, the prediction on the hadronic
branching ratios, $R_c$, etc..., can be made once we extract
$g_{new,q}$ from $R_b$. (The actual values of $g_{new,q}$
will be given below in the discussion of the compositeness
scale.) For branching ratios and other quantities 
involving leptons, one needs to know $g_{new,l}$. One
can, for instance, choose the range of $g_{new,l}$ so as to fit
$R \equiv \Gamma(had) / \Gamma(l {\bar l})$ and predict
what other quantities such as $\Gamma_Z$, $A_{LR}$, and
$\sigma_{had}$ might be. This is the procedure that we shall
follow below. As we shall see, it turns out that the range of
parameters that fits $R$
will predict $A_{LR}$ to be consistent with the SLD data rather
than the corresponding LEP data.

We list in Table 3 the predicted values for $R$, $A_{LR}$,
$\Gamma_Z$, and $\sigma_{had}$ for the range of
$g_{new,l} = 0.02-0.035$. (The value of $\alpha_S (M_Z)$
used in this paper is 0.125.)
The range for $g_{new,l}$ is chosen so as to
show the correlation between $R$ and the other quantities.
Let us first notice the following behaviour. $R$ ($A_{LR}$)
increases (decreases) as $g_{new,l}$ decreases from 0.035
to 0.02. What happens when $g_{new,l}$ is less than 0.02?
Although not listed in  the table, it turns out that, for $g_{new,l} =
0.01$, the place where $A_{LR}$ is $1\:\sigma$ from the
the SM prediction and the LEP result, namely $A_{LR} = 0.148$
gives a value to $R$ (= $21 \pm 0.09$) which is at least
$4\:\sigma$ away from the experimental value ($20.788 \pm 0.032$).
If we choose $g_{new,l} \stackrel{>}{\sim} 0.02$ 
so that the predicted $R$ agrees with the experimental values,
one can see from Table 3 that $A_{LR}$ is predicted to be
more in agreement with the SLD data than the LEP data.
It means that our model cannot satisfy all of the LEP data.
The discrepancy between the LEP and SLD data for $A_{LR}$
might have pointed toward some kind of new physics such as
the one described in this manuscript. Needless to say, it is
important to resolve this discrepancy in order to be able to make
any kind of statement concerning new physics in this sector.

What is the meaning of the range $g_{new,l} =0.02-0.035$?
Since $g_{new,q}$ is fixed by $R_b$, the remaining free
parameter is $g_{new,l}$. But it is itself constrained by
the other electroweak observables (Z width, etc...) which are
basically consistent with the standard model predictions
except for the SLD measurement of $A_{LR}$.
It is precisely because of these
features that the above range of $g_{new,l}$ is chosen
so as to be consistent with these observables. We shall come
back again to this point below.

Another observation can be made by looking at Table 3. One
notices that both $R$ and $\Gamma_Z$ increase with
increasing resonance mass $M_V$. For $M_V \geq 96$ GeV,
the predicted $\Gamma_Z$ will be at least 2 $\sigma$ away
from the fairly precise experimental value of $2.4963
\pm 0.0032$ GeV and this worsens as $g_{new,l}$ gets smaller.

The predictions for $R$, $\Gamma_Z$, $A_{LR}$, and $\sigma_{had}$
depend on the Higgs mass, although not in a significant way. Table
3 presents predictions where $m_H$ is taken to be 700 GeV. As we
lower the Higgs mass to 100 GeV, there is an increase in these
predictions by approximately 0.2 \%. So these predictions are
not very sensitive to the Higgs mass.

Let us now summarize our results. Table 2 lists the predictions
for $R_c$ and $R_s$ as a function of $m_t$. In particular
we notice the decrease in $R_c$. Table 3 lists the predictions
for $R$, $A_{LR}$, $\Gamma_Z$ and $\sigma_{had}$ as a function
of $g_{new,l}$, $m_t$ and $M_V$. There we notice that, by
fixing $R$ to agree with the experimental values, our
prediction for $A_{LR}$ tends to agree with the SLD result.
In addition, the preferred range for the new resonance mass
is between 92.5 GeV and 96 GeV. Translated into ${\cal P}$ mass,
the range is between $m_{\cal P}$ = 47 GeV to $m_{\cal P}$ = 48.8
GeV.

The above results, namely $M_V = 92.5-96$ GeV, trigger the
obvious question: Is such mass range already ruled out by experiment?
The answer is negative. The reasons are twofold.

First, let us 
estimate the V width. Since $M_V < m_t$, V decay into
five quark flavors and six lepton flavors. Its width is then
given by $\Gamma_V = (1/12 \pi)(15 g_{new,q}^2 +6 g_{new,l}^2)
M_V$. To illustrate our point, let us take $M_V = 96$ GeV
and $m_t = 170$ GeV. From $R_b$, we extract $g_{new,q} =
0.096-0.163$ (the range comes from the spread in $R_b$).
Let us take the maximum allowed value for $g_{new,l}$,
namely 0.035. The V width is then estimated to be
$\Gamma_V = 0.37-1.03$ GeV. In the LEP 1990-1991 run,
the energy scan was $\sqrt{s} = M_Z \pm 3 GeV$. The 1995
run has an energy scan $\sqrt{s} = 130- 140 GeV$. In
consequence, V with a mass 96 GeV would not have been seen
{\em directly}. In fact one can safely say that $M_V=
94.5 - 96$ GeV would be outside the range of direct detection.
The lower mass range, 92.5 GeV - 94.5 GeV, is more problematic
although it is possible that one might miss
such V in that mass range. In any case, at least the range
94.5 GeV - 96 GeV does not appear to be ruled by the present
energy scan.

The second point concerns the upper bound of 96 GeV. This
value is by no means a firm prediction. It depends on a number
of things: the experimental spread of the electroweak observables,
the spread in $\alpha_S$, etc...This upper bound could easily
be higher than 96 GeV by a few GeV. An extensive analysis
will be carried out in a forthcoming article.

As far as direct detection is concerned, the model is well
and alive. We would like to suggest  a LEP scan of 100 GeV
down to $M_Z$. It will be a crucial test of this model.

Have we or have we not seen V indirectly through electroweak
observables away from the V resonance? The most obvious places
to look at are the Z width and $\sigma_{had}$. As we have discussed
above and as shown in Table 3, as long as $g_{new,l} = 0.02-
0.035$, our numbers for these quantities agree with the experimental
numbers, which themselves are consistent with the Standard Model. So by
just looking at $\Gamma_Z$ and $\sigma_{had}$, one {\em cannot}
tell whether V is there or not. Since part of the motivation
for building this model was to explain the {\em discrepancy} between the
experimental results for $R_b$ and $R_c$ and the SM predictions,
would such a discrepancy be an indirect manifestation of V?
Furthermore, as we have mentioned earlier, there is a discrepancy
between the SLD result for $A_{LR}$ and that coming from LEP
as well as the SM prediction. Since our result agrees with the
SLD one, would that again be an indirect manifestation of V?

To summarize, there is yet no direct nor indirect evidence
against our model. On the contrary, there might already be
some indirect evidence for some new phenomena of the types
described here. Again an energy scan from approximately
100 GeV down to $M_Z$ is a crucial test for our model.

Let us now turn to the other members of this {\em nonsequential}
family, the ${\cal R}$ quark and the leptons ${\cal N}$ and
${\cal E}$. What constraints can one obtain on the masses
of these particles? One obvious constraint is the fact that
they have to be {\em heavier} than $M_Z /2$ since they have not
yet been seen.

To be able to say more
than this, one has to invoke additional information.
This is where the S and T parameters \cite{peskin},
or the $\epsilon$ parameters of \cite{altarelli}, come in.
To be able to use these parameters in our context, one has to have
an effective $SU(2)_L \otimes U(1)_{Y}$ theory. This is possible
if the extra $Z^{\prime}$ mixes very little with the SM $Z$.
At the beginning of the section on the phenomenological
analysis of $Z \rightarrow b {\bar b}$, we have discussed this
possibility and we have referred to an analysis done by Ref.
\cite{altarelli,luo} concerning electroweak precision constraints
on extended gauge models such as the one considered here. There
it was found that the mixing angle between $Z$ and $Z^{\prime}$
is constrained to be less than 1 %. Also in Ref.
\cite{altarelli}, an explicit contribution of $Z^{\prime}$
to the $\epsilon$ parameters was given. It can be
seen there that this contribution is negligibly small
for very small mixing and one is practically back to
the SM analysis. We refer the reader to Ref. \cite{altarelli,luo}
for more details. In consequence, we
shall {\em assume} in this paper that this mixing, 
which depends on the details
of the Higgs sector, is negligible (less than at most 1 \%) and
its effects on electroweak precision measurements such
as the oblique parameters S and T
can be neglected. In consequence, one has practically an
effective $SU(2)_L \otimes U(1)_{Y}$ theory. We can then make use of
the most up-to-date determination of S and T to constrain the
masses of ${\cal R}$, ${\cal N}$, and ${\cal E}$.

Before carrying out this analysis, a useful remark is in
order here.
Since this new family is {\em nonsequential}, there is no reason
to expect the masses and mass splitting (between up and down members) 
to be "similar" in pattern to the other three families.

We use the most recent determination of the new physics
contribution to S and T as fitted by
Ref. \cite{langacker}. They are:
\begin{mathletters}
\begin{equation}
S_{new} = -0.28 \pm 0.19 \stackrel{-0.08}{+0.17},
\end{equation}
\begin{equation}
T_{new} = -0.20 \pm 0.26 \stackrel{+0.17}{-0.12}.
\end{equation}
\end{mathletters}.

We shall use $S_{new}^{max} = 0.08$ and $T_{new}^{max} =0.23$.

To compute $S_{new}$ and $T_{new}$ in our model, we need,
in addition to the range for $m_{\cal P}$ quoted above,
one more input: the mass of one lepton which we shall
choose to be the mass of the heavy neutrino. Since this is not meant
to be an exhaustive discussion, we shall restrict ourselves to
a neutrino mass of 46 GeV (other starting values will be included
in a more comprehensive analysis).
We shall assume  
that the neutrino, ${\cal N}$, is
a Majorana particle. The contribution of the leptons
to S and T can now easily be computed \cite{gates}. 

As mentioned above, we now require (from $S_{new}^{max} = 0.08$ and
$T_{new}^{max} = 0.23$)
\begin{mathletters}
\begin{equation}
S_{new}^{q} + S_{new}^{l} \leq 0.08,
\end{equation}
\begin{equation}
T_{new}^{q} + T_{new}^{l} \leq 0.23. 
\end{equation}
\end{mathletters}
For a given
$m_{\cal P}$ and $m_{\cal N}$, we compute S and T for a range of
$m_{\cal R}$ and $m_{\cal E}$, keeping in mind the above constraint.
We now list the relevant values, all of them computed with
$m_{\cal N} = 46$ GeV.

For $m_{\cal P} = 47 $ GeV, we have: 1) $S_{new}^{q} = 0.122$,
$S_{new}^{l} = -0.042$, $T_{new}^{q} = 0.022$,
$T_{new}^{l} = 0.208$ corresponding to $m_{\cal R} = 67$ GeV
and $m_{\cal E} = 162$ GeV; 2) $S_{new}^{q} = 0.072$,
$S_{new}^{l} = 0.008$, $T_{new}^{q} = 0.195$,
$T_{new}^{l} = 0.035$ corresponding to $m_{\cal R} = 107$ GeV
and $m_{\cal E} = 97$ GeV. Notice that as $m_{\cal R}$
increases, $m_{\cal E}$ decreases. The allowed ranges
for $m_{\cal R}$ and $m_{\cal E}$ are therefore
$67 GeV \leq m_{\cal R} \leq 107 GeV$ and
$162 GeV \geq m_{\cal E} \geq 97 GeV$. Any other value
outside that range is incompatible with the constraint.

For $m_{\cal P} = 48.8 $ GeV, we have: 1) $S_{new}^{q} = 0.122$,
$S_{new}^{l} = -0.042$, $T_{new}^{q} = 0.023$,
$T_{new}^{l} = 0.207$ corresponding to $m_{\cal R} = 69$ GeV
and $m_{\cal E} = 162$ GeV; 2) $S_{new}^{q} = 0.078$,
$S_{new}^{l} = 0.002$, $T_{new}^{q} = 0.172$,
$T_{new}^{l} = 0.058$ corresponding to $m_{\cal R} = 105$ GeV
and $m_{\cal E} = 110$ GeV. The allowed ranges are
$69 GeV \leq m_{\cal R} \leq 105 GeV$ and
$162 GeV \geq m_{\cal E} \geq 110 GeV$.

Again the above results refer to $m_{\cal N} = 46 GeV$. 
For $m_{\cal N} = 48 GeV$, the ranges are slightly
modified (the lower bounds are slightly
higher). A more comprehensive analysis for various values
of $m_{\cal N}$ will be
presented elsewhere. 

What are the implications of the above constraints on $m_{\cal R}$
and $m_{\cal E}$ coming from S and T?

First, $m_{\cal E}$ has to be at least 97 GeV, and most likely
at least 110 GeV. Therefore ${\cal E}$
is not likely to be found at LEP2. What could
be found at LEP2 would be at least {\em one} new threshold, the
${\cal P}$ quark, and possibly two, the ${\cal R}$ quark, if it is
light enough. What we mean by new threshold here is simply the
appearance of the first resonance (lowest lying $Q {\bar Q}$
state). 
It would be an experimental challenge to find the
nonsequential charged lepton with mass greater than 97 or 110 GeV at
hadron colliders. One interesting scenario is when both quarks
might be found at LEP2, e.g. when $m_{\cal P} = 49$ GeV,
$m_{\cal R} = 81$ GeV. The constraint from S and T would imply that
that $m_{\cal E} \sim 136$ GeV. How would one detect such a heavy
nonsequential charged lepton?

As alluded to in the beginning of the paper, this new
family naturally involves new physics which can give rise
to non-standard decays of the ${\cal R}$ quark and consequently
invalidates the CDF and D0 limits of 118 GeV and 131 GeV. In
such a case, the lower limit on the ${\cal R}$ quark would
be 62 GeV. Our own lower limits on the ${\cal R}$ quark
mass are higher than that value. Now, from Eq.\ (\ref{LB}) it follows that if
$g_{B}^2/\Lambda^2 \geq g^2/2 M_W^2$, where $g$ is the weak
coupling, the leptonic decay of ${\cal R}$ would be mostly into ${\cal P}
\tau \nu$. Below we shall see if it is reasonable to have
such a constraint.
This opens up
the possibility that the whole new quark family can be found by LEP2.
First the R ratio would be 16/3 or at least 12/3 = 4 (if the
${\cal R}$ quark mass is above 81 GeV). Some words of caution
are in order. The last number depends
of course on being able to detect the ${\cal P}$ quark which
could conceivably escape the detector because of its
possible relatively "long" lifetime. In that case, the R ratio
would probably be unchanged, namely 11/3, i.e. one would not see
the ${\cal P}$ quark even if one were above its open threshold.
The 1995 run of LEP1.5 with a center of mass
energy well above the open ${\cal P}$ threshold,
did not show any increase in the R ratio. It implies that the
${\cal P}$ lifetime should be long enough for it to escape the
detector. This is conceivable in our scenario since ${\cal P}$
is a {\em nonsequential} quark with little mixing to the
other three families and, consequently, could have a long lifetime.
In some sense, the LEP1.5 result puts a constraint on the minimum
lifetime ${\cal P}$ could have. In fact, one can put a rough limit
on the mixing of ${\cal P}$ with the other quarks using the
LEP1.5 constraint. One can compute the mean decay length of
${\cal P}$ (see e.g. \cite{barger} on p. 76) taking $\sqrt{s} =
130$ GeV, $m_{{\cal P}} = 47$ GeV. Requiring the decay length
to be approximately greater than say 10 m, one finds that
$\|V \|^{2} = \| V_{Pc} \|^{2} + \| V_{Pu} \|^{2}$ should be
less than $3 \times 10^{-13}$ giving $\|V\| \leq 5 \times 10^{-7}$.
Needless to say, this is only a rough limit.
Incidentally, this limit is consistent with the cosmologically
comment made below. The ${\cal P}$ quark can form neutral and
charged mesons with the light quarks. One might wonder if
the charged mesons might not leave some tracks in the detector.
This is an experimental issue which needs to be carefully examined
to see if these kinds of tracks might have been missed.
If ${\cal R}$ is light enough
(below 80 GeV), it could be produced at LEP2 but its identification
might be tricky since its main decay is ${\cal R} \rightarrow
{\cal P} f \bar{f}$, where $f$ is a standard light fermion,
and ${\cal P}$ can escape the detector. In any case, it would be
interesting to watch out for unusual events related to the one
just mentioned. If ${\cal R}$ is heavier than 81 GeV then one
would not see any increase in the R ratio even at LEP2 since
there was none at LEP1.5. In this case the direct detection
of this new, nonsequential family of quarks will have to
rely on hadron machines.

As we have emphasized earlier, by {\em nonsequential} we
really mean that there is very little mixing of this
new family with the other three. This tiny mixing would be
enough to evade cosmological constraints on stable quarks
\cite{nardi}.
Even with a mixing as small as, e.g. ${\lambda}^{10}$ between ${\cal P}$
and the charmed quark, the ${\cal P}$ lifetime would be of the
order $10^{-8}$ sec which is certainly fast enough to evade
any of such constraint.
 
Finally we would like to say a few words about the scales of
"compositeness" in our model.

A four-fermi coupling as given by ${\cal L}_{q0}$ would be
diagramatically similar to a quark diagram for meson-meson
scattering except that here we would have a scalar line
instead of one of the two quark lines. It follows that
$g_{new,q}$ is not necessarily given by the wave function at
the origin. We shall {\em assume}, for the sake of estimate, that we can 
write 
\begin{equation}
g_{new,q} \equiv (g_{q}^2/\Lambda^2)g_H^2 F_V, 
\end{equation}
where $g_H^2$ represents the rescattering of the scalar components.
If $g_q^2/ 4 \pi=g_H^2/ 4 \pi= 2.5$, $\Lambda$ can be computed 
in terms of $R_b$ using the definition of $g_{new,q}$ discussed
above. Under the above dynamical assumption, the values for
$\Lambda$ are listed in Table 5. The range of $\Lambda$ for
each value of $M_V$ corresponds to $R_{b}^{max}$ and $R_{b}^{min}$.
Notice that these values can easily be
underestimated by a factor of two or so. The point is
that they do not have to be as high as 10 or 100 TeV.
In summary the scale of "compositeness" in our model can be as low
as a few TeVs. Caution should be applied to the literal
interpretation of $\Lambda$ as the "compositeness" scale which,
in general, might not be too different from $\Lambda$ itself.

Are these estimates consistent with experiment? 
Is there any "evidence" for compositeness? We shall briefly
address these questions below.

One word of caution
is in order here. Present experiments probing compositeness only
deal with operators of the type represented in Eq.\ (\ref{Lf})
which involve only fermions of the first three generations. As
we have stated above, the coefficients $g_i^2/\Lambda_i^2$ are
not necessarily identical to $g_q^2/\Lambda^2$. The CDF limit
on quark "compositeness" based on dijet invariant mass spectrum
set a limit of $\Lambda_i \geq 1.4 TeV$ for $g_i^2 = 4 \pi$.
(This is also consistent with the TRISTAN limit.)
There is some indication that the CDF high $p_T$ data
shows some discrepancy with the QCD prediction. Much work needs
to be done in order to clarify both the experimental situation and
the QCD prediction (gluon distribution, etc...), but there is
a possibility that it is a signal for quark compositeness
with a "low" compositeness scale. We shall see in the not-too-distant
future whether this possibility is true or not.
{\em Even if} we assume that the above coefficients are similar then, 
taking into account the uncertainty in extracting $\Lambda$ from $g_{new,q}$
described above, we can safely say that our crude estimate is not
inconsistent with the experimental lower bound. Needless to say,
much more detailed studies are needed to lay out the various
constraints on dynamical assumptions coming from experiment.
Some of these issues will be dealt with in a subsequent work.

If $g_B^2/\Lambda_B^2$ as written in Eq.\ (\ref{LB}) were
similar to $g_q^2/\Lambda^2$ then the characteristic
strength would be approximately $ 2 \cdot 10^{-5}$ $GeV^{-2}$. 
This is to be compared with a characteristic weak interaction strength
$g^2/2 M_W^2$ at the ${\cal R}$ mass which is approximately
$3 \cdot 10^{-5}$ $GeV^{-2}$. Even with the above assumption, one 
can see that the leptonic decay of ${\cal R}$ will be mainly in the
channel ${\cal P} \tau \nu$. Relaxing that assumption can
make this mode even stronger if $g_B^2/\Lambda_B^2$ is given
a larger value. The CDF and D0 limits of 118 GeV and 131 GeV
assuming standard decay will no longer be applicable. The
lower limit of 62 GeV will then be applicable. 
This is consistent with our lower bound of approximately
67 GeV on the mass of ${\cal R}$.

\section{Conclusion} 

We have presented a simple scenario to explain the "anomaly"
in $R_b$ and, as a consequence, we have made a number of
predictions, $R_c$, etc..., including the presence of a new, {\em non-
sequential} fourth family, some of  whose members
could have masses below
$M_W$, an exciting prospect for near-future discoveries. In
particular, a charge -1/3 quark is predicted to lie between
47 and 49 GeV. Its charge +2/3 companion is constrained
to lie above 67 GeV and, as a consequence, it is possible to have
two (but {\em at least} one) new thresholds below $M_W$. An energy
scan from 100 GeV down to $M_Z$ would provide a crucial test
of this model.
With the
constraint that the nonsequential neutrino be heavier than $M_Z/2$,
the nonsequential charged lepton turned out to be heavier than 97 GeV.
These predictions
are firm and the model can be easily disproved if none of these
particles are found within the capability
of LEP2. Since this is a composite model,
we have estimated the "compositeness" scale to be "low", i.e.
below approximately 5 TeV. This is relevant to a suggestion (to
be confirmed) that signals of "compositeness" might have been
seen at CDF.

This work was supported in part by the U.S. Department of Energy
under grant No. DE-A505-89ER40518.

% now the references. delete or change fake bibitem. delete next three
%   lines and directly read in your .bbl file if you use bibtex.

%  that you will use with \ref{} command in the braces of the \label{} command.
%

\begin{table}
\caption{The Standard Model predictions for $R_{f}^{SM} \equiv
\Gamma(Z \rightarrow f \bar{f}) / \Gamma(had)$ as functions of the
Higgs boson and top quark masses. Here
$R_{e}^{SM} \equiv \Gamma(had) / \Gamma(e {\bar e})$}
\begin{tabular}{clclclclclclcl}
\multicolumn{1}{c}{$m_H(GeV)$} &\multicolumn{1}{c}{$m_t(GeV)$} 
&\multicolumn{1}{c}{$R_{b}^{SM}$} 
&\multicolumn{1}{c}{$R_{c}^{SM}$}
&\multicolumn{1}{c}{$R_{s}^{SM}$}
&\multicolumn{1}{c}{$R_{u}^{SM}$}
&\multicolumn{1}{c}{$R_{e}^{SM}$} \\
\tableline
100  &150  &0.2162  &0.172  &0.2199  &0.172  &20.7818  \\
100  &160  &0.21586  &0.172  &0.22  &0.172  &20.779   \\
100  &170  &0.2155  &0.1722  &0.22  &0.1722  &20.776  \\
100  &190  &0.21473  &0.1724  &0.22  &0.1725  &20.769  \\
700  &150  &0.2162  &0.172  &0.22  &0.172  &20.763  \\
700  &160  &0.2159  &0.172  &0.22  &0.172  &20.76  \\
700  &170  &0.21554  &0.1721  &0.22  &0.1721  &20.758  \\
700  &190  &0.21477  &0.1723  &0.22  &0.1724  &20.752  \\
\end{tabular}
\end{table}
\begin{table}
\caption{Predictions for the ratios $R_c=R_u$, $R_s=R_d$,
as functions of $m_t$}
\begin{tabular}{clclcl}
\multicolumn{1}{c}{$m_t(GeV)$} 
&\multicolumn{1}{c}{$R_{c}$}
&\multicolumn{1}{c}{$R_{s}$} 
& \multicolumn{1}{c}{$R_{c}^{exp}$} \\
\tableline
150  &0.1634 $\mp$ 0.0026  &0.2258 $\pm$ 0.0017 &0.1540 $\pm$ 0.0074   \\
160  &0.1629 $\mp$ 0.0026  &0.2261 $\pm$ 0.0017 &0.1540 $\pm$ 0.0074  \\
170  &0.1624 $\mp$ 0.0026  &0.2265 $\pm$ 0.0017 &0.1540 $\pm$ 0.0074  \\
190  &0.1615 $\mp$ 0.0026  &0.2273 $\pm$ 0.0017 &0.1540 $\pm$ 0.0074  \\
\end{tabular}
\end{table}
\begin{table}
\caption{Predictions for  $R\equiv \Gamma(had) / \Gamma(l {\bar l})$, 
$A_{LR}$, $\Gamma_Z$, and $\sigma_{had}$ 
as functions of $M_V$, $m_t$ and $g_{new,l}$. They are to be
compared with the following experimental values: $R=20.788 \pm 0.032$,
$A_{LR}(SLD)= 0.1551 \pm 0.004$, $A_{LR}(LEP)= 0.139 \pm 0.0089$,
$\Gamma_Z(GeV)= 2.4963 \pm 0.032$, and $\sigma_{had}(nb)=41.488 \pm 0.078$.
The two values for each prediction correspond to $g_{new,l}$ =
0.02, 0.035 respectively.}
\begin{tabular}{clclclclcl}
\multicolumn{1}{c}{$M_V,m_t$(GeV)} 
&\multicolumn{1}{c}{$R$}
&\multicolumn{1}{c}{$A_{LR}$} 
& \multicolumn{1}{c}{$\Gamma_Z(GeV)$} 
&\multicolumn{1}{c}{$\sigma_{had}(nb)$} \\
\tableline
92.5,150  &(20.8,20.6) $\pm$ 0.09,  &0.155,0.162 
&(2.507,2.502) $\pm$ 0.008 &(41.91,42.45) $\pm$ 0.44   \\
92.5,160  &(20.817,20.623) $\pm$ 0.09  &0.155,0.162 
&(2.511,2.508) $\pm$ 0.008 &(41.9,42.44) $\pm$ 0.44 \\
92.5,170  &(20.833,20.64) $\pm$ 0.09  &0.155,0.162 
&(2.516,2.512) $\pm$ 0.008 &(41.88,42.42) $\pm$ 0.44 \\
92.5,190  &(20.87,20.67) $\pm$ 0.09  &0.155,0.162 
&(2.527,2.522) $\pm$ 0.008 &(41.86,42.4) $\pm$ 0.44 \\
94,150  &(20.867,20.72) $\pm$ 0.09,  &0.153,0.158 
&(2.508,2.505) $\pm$ 0.008 &(41.74,42.14) $\pm$ 0.44   \\
94,160  &(20.88,20.733) $\pm$ 0.09  &0.153,0.158 
&(2.511,2.508) $\pm$ 0.008 &(41.9,42.44) $\pm$ 0.44 \\
94,170  &(20.9,20.8) $\pm$ 0.09  &0.153,0.158 
&(2.518,2.514) $\pm$ 0.008 &(41.71,42.12) $\pm$ 0.44 \\
94,190  &(20.93,20.78) $\pm$ 0.09  &0.153,0.158 
&(2.528,2.525) $\pm$ 0.008 &(41.68,42.1) $\pm$ 0.44 \\
96,150  &(20.93,20.83) $\pm$ 0.09,  &0.15,0.153 
&(2.51,2.508) $\pm$ 0.008 &(41.56,41.83) $\pm$ 0.44   \\
96,160  &(20.95,20.85) $\pm$ 0.09  &0.15,0.153 
&(2.514,2.512) $\pm$ 0.008 &(41.55,41.82) $\pm$ 0.44 \\
96,170  &(20.96,20.86) $\pm$ 0.09  &0.15,0.153 
&(2.52,2.517) $\pm$ 0.008 &(41.54,41.8) $\pm$ 0.44 \\
96,190  &(21.,20.9) $\pm$ 0.09  &0.15,0.153 
&(2.53,2.527) $\pm$ 0.008 &(41.51,41.78) $\pm$ 0.44 \\
\end{tabular}
\end{table}
\begin{table}
\caption{The values of the "compositeness" scale $\Lambda$ as a function
of $M_V$= 92.5-96 GeV and $m_t$= 150-190 GeV. The spread reflects the error
in $R_b$.}
% \label{}
\begin{tabular}{clclclclcl}
\multicolumn{1}{c}{$\Lambda$(TeV)} 
&\multicolumn{1}{c}{150} 
&\multicolumn{1}{c}{160}
&\multicolumn{1}{c}{170} 
& \multicolumn{1}{c}{190} \\
\tableline
92.5 &1.9-2.51 &1.87-2.42 &1.83-2.34 &1.76-2.17 \\
94   &1.64-2.2 &1.61-2.12 &1.58-2.04 &1.51-1.89 \\
96   &1.36-1.83 &1.33-1.76 &1.3-1.7 &1.24-1.57  \\
\end{tabular}
\end{table}

\end{document}